\documentstyle[12pt]{article}
\def\zid{1\kern-0.36em\llap~1}

\newcommand{\beq}{\begin{equation}}
\newcommand{\ber}{\begin{eqnarray}}
\newcommand{\eeq}{\end{equation}}
\newcommand{\eer}{\end{eqnarray}}

\begin{document}

\begin{titlepage}
\rightline{[SUNY BING 5/21/01R3 ] }  \rightline{ hep-ph/010638}
\vspace{2mm}
\begin{center}
{\bf \hspace{1.85 cm} THREE NUMERICAL PUZZLES \newline AND THE TOP
QUARK'S CHIRAL WEAK-MOMENT  }\\ \vspace{2mm} Charles A.
Nelson\footnote{Electronic address: cnelson @ binghamton.edu  } \\
{\it Department of Physics, State University of New York at
Binghamton\\ Binghamton, N.Y. 13902-6016}\\[2mm]
\end{center}


\begin{abstract}
Versus the standard model's $t \rightarrow W^{+} b $ decay
helicity amplitudes, three numerical puzzles occur at the $0.1\%$
level when one considers the amplitudes in the case of an
additional $``(f_M + f_E)"$ coupling of relative strength
$\Lambda_{+} \sim 53 GeV$. The puzzles are theoretical ones which
involve the $t \rightarrow W^{+} b $ decay helicity amplitudes in
the two cases, the relative strength of the additional $(f_M +
f_E)$ coupling, and the observed masses of these three particles.
A deeper analytic realization is obtained for two of the puzzles.
Equivalent analytic realizations are given for the remaining one.
An empirical consequence of these analytic realizations is that it
is important to search for effects of a large chiral weak-moment
of the top-quark, $\Lambda_{+} \sim 53 GeV$.  A full theoretical
resolution would include relating the origin of such a chiral
weak-moment and the mass generation of the top-quark, the W-boson,
and probably the b-quark.
\end{abstract}

\end{titlepage}

\section{Introduction}

Versus the standard model's (SM) $t \rightarrow W^{+} b $ decay
helicity amplitudes, three numerical puzzles occur at the $0.1\%$
level when one considers the amplitudes in the case of an
additional $``(f_M + f_E)"$ coupling of relative strength
$\Lambda_{+} \sim 53 GeV$ [1].  The puzzles are theoretical ones
which involve the $t \rightarrow W^{+} b $ decay helicity
amplitudes in the two cases, the relative strength of the
additional $(f_M + f_E)$ coupling, and the observed masses of
these three particles (see discussion at end of this Section).
While the observed mass-values of the top-quark, W-boson, and
b-quark are involved here, it is not a matter of presently
available empirical data disagreeing with a SM prediction.  The
puzzles arose in a search for empirical ambiguities between the
standard model's $(V-A)$ coupling and possible single additional
Lorentz couplings that could occur in the ongoing [2] and
forthcoming [3,4] top-quark decay experiments at hadron and $l^-
l^+$ colliders.

In this paper we report analytic realizations of these three
numerical puzzles. There are four types of analytical relations
which are listed below as (i)-(iv) in Section 2.

For \hskip1em $t \rightarrow W^+ b$, the most general Lorentz
coupling is $ W_\mu ^{*} J_{\bar b t}^\mu = W_\mu ^{*}\bar
u_{b}\left( p\right) \Gamma ^\mu u_t \left( k\right) $ where $k_t
=q_W +p_b $, and
\begin{eqnarray}
\Gamma _V^\mu =g_V\gamma ^\mu + \frac{f_M}{2\Lambda }\iota \sigma
^{\mu \nu }(k-p)_\nu + \frac{g_{S^{-}}}{2\Lambda }(k-p)^\mu
\nonumber \\ +\frac{g_S}{2\Lambda }(k+p)^\mu
+%
\frac{g_{T^{+}}}{2\Lambda }\iota \sigma ^{\mu \nu }(k+p)_\nu
\end{eqnarray}
\begin{eqnarray}
\Gamma _A^\mu =g_A\gamma ^\mu \gamma _5+ \frac{f_E}{2\Lambda
}\iota \sigma ^{\mu \nu }(k-p)_\nu \gamma _5 +
\frac{g_{P^{-}}}{2\Lambda }(k-p)^\mu \gamma _5  \nonumber \\
+\frac{g_P}{2\Lambda }%
(k+p)^\mu \gamma _5  +\frac{g_{T_5^{+}}}{2\Lambda }\iota \sigma
^{\mu \nu }(k+p)_\nu \gamma _5
\end{eqnarray}
For a most general treatment of additional Lorentz structures to
pure $V-A$, the $g_i$ or $\Lambda_i$ must be considered as complex
phenomenological parameters.

To avoid confusion, two notational details should be noted in our
usage of Eqs. (1,2) in this paper:  First, we will more often use
chiral combinations of the couplings which are defined by $g_{L,R}
\equiv g_V \mp g_A,$ $g_{\pm} \equiv g_{f_M \pm f_E} \equiv f_M
\pm f_E$, $g_{S \pm P} = g_S \pm g_P, \ldots $ (for others see
appendix). Second, to avoid cumbersome notation and verbiage, we
find it very convenient to suppress the analogous subscript on the
effective mass-scales, $\Lambda_{i}$. For example, in (1) a reader
might choose to write $ \frac{f_M}{2\Lambda } $ as
$\frac{f_M}{2\Lambda_{M} }$.  However, a smarter reader would
realize that in the analysis in this paper for each $i$ value only
the ratio of $g_i$ and $\Lambda_{i}$ ever occurs so one could just
omit the $\Lambda_{i}$ by completely absorbing it into its $g_i$.
Nevertheless, we proceed a little differently: we use the
unsubscripted $\Lambda$'s in (1,2) which does keep track of the
mass-dimensions and then we subscript them when we must. For
instance, whenever we use ``$g_L = 1$ units with $g_i = 1$", we do
move the corresponding subscript from the $g_i$ to its
$\Lambda_i$.  In essence, this is but a slight variation on the
smarter reader's procedure but one which, we believe, is
worthwhile in this paper. For present purposes, referring to a
$\Lambda_i$ scale in ``$g_L = 1$ units with $g_i = 1$" is a simple
way to characterize the size and sign of the additional effective
$ \frac{g_i}{2\Lambda_i } $ coupling. [ When a fundamental theory
leads to a specific coupling in (1,2), both $f_M$ and $\Lambda_M$,
and not just their ratio, would normally be independently defined
and observable.  In general, one would then not want to omit
either such quantity nor any such subscripts!  A more fundamental
or more general realization of electroweak symmetry in the context
of the SM might also yield deviations from $g_L =1, g_i = 0$ limit
but we do not consider that possibility here. ]

As discussed in greater detail in the earliest paper in [1],  the
various couplings in (1,2) are not Lorentz independent.  In
particular, to obtain the identical $\lambda_b= - \frac{1}{2}$
amplitudes as for the pure $V-A$ coupling of the SM, there is a
dominant scale of $\Lambda_{S+P} = - \Lambda_{f_M + f_E } \sim
\frac{m_t}{2} = 87GeV$ and there is a negligible (for $\lambda_b=
- \frac{1}{2}$ amplitudes) scale $\Lambda_{S-P} = - \Lambda_{f_M -
f_E } \sim - \frac{ (m_t)^2}{2 m _b } = -3,400GeV$. These scales
are, of course, not to be confused with those for the 2
``dynamical phase-type ambiguities" [1] discussed below in this
Section in association with Table 1; the ``ambiguity scale" for
the additional Lorentz structure is respectively $\Lambda_{S + P}
\sim -35 GeV$, $\Lambda_{f_M + f_E} \sim 53 GeV$.

These various coupling scales can be compared with the SM's EW
scale $v_{EW}=\sqrt{-\mu^2 / \vert \lambda \vert} = \sqrt{2}
\langle 0|\phi |0\rangle \sim 246GeV$, where $\phi$ is the Higgs
field. These scales can also be compared with the ``few TeV"
cut-off scale needed to control the one-loop quadratic-divergent
radiative corrections which arise in the Higgs mass
renormalization in the SM; this ``few TeV" scale thereby occurs in
the SUSY and technicolor generalizations of the SM.

In the  $t$ rest frame, the matrix element for $t \rightarrow
W^{+} b$ is
\begin{equation}
\langle \theta _1^t ,\phi _1^t ,\lambda _{W^{+} } ,\lambda _b
|\frac 12,\lambda _1\rangle =D_{\lambda _1,\mu }^{(1/2)*}(\phi
_1^t ,\theta _1^t ,0)A\left( \lambda _{W^{+} } ,\lambda _b \right)
\end{equation}
where $\mu =\lambda _{W^{+} } -\lambda _b $ in terms of the $W^+$
and $b$-quark helicities.  The helicity amplitudes, $A\left(
\lambda _{W^{+} } ,\lambda _b \right)$, corresponding to the
couplings in (1) and (2) are listed in the Appendix.  We will
denote respectively the ``Standard Model's" and the ``$(V-A) +
(f_M + f_E)$" coupling's amplitudes by an (SM) and (+) subscript.

In (3), the asterisk denotes complex conjugation.  The final
$W^{+}$ momentum is in the $\theta _1^t ,\phi _1^t$ direction and
the $b$-quark momentum is in the opposite direction. $\lambda_1$
gives the $t$-quark's spin component quantized along the $z$ axis.
$\lambda_1$ is also the helicity of the $t$-quark if one has
boosted, along the ``$-z$" direction, back to the $t$ rest frame
from the $(t \bar{t})_{cm}$ frame.  It is this boost which defines
the $z$ axis in the $t$-quark rest frame for angular analysis [1].

To be able to quantitatively assess future measurements of
competing observables in $t \rightarrow W^+ b $ decay, we
considered in [1] the $g_{V-A}$ coupling values of helicity decay
parameters versus those for `` $(V-A)$ $+$ single additional
Lorentz structures."  Versus the SM's dominant L-handed $b$-quark
amplitudes, there are 2 dynamical phase-type ambiguities produced
respectively by an additional $(S+P)$ and by an additional $(f_M +
f_E)$ coupling, see the $ A \left( 0,-\frac 12\right) $ and $ A
\left( -1,-\frac 12\right) $ columns of Table 1. For a reader
interested in more details about this table, we note that in the
earliest reference in [1], this same table is discussed at length
in association with another table with the corresponding
consequences for the observable helicity parameters for $t
\rightarrow W^+ b$ decay in the cases that the observed top-quark
is coupled respectively by ``$(V-A) + (S + P) $" and ``$(V-A) +
(f_M + f_E) $".  By tuning the effective-mass-scale associated
with the additional coupling constant, the additional $(S+P)$
coupling, $(f_M + f_E)$ coupling, has respectively changed the
sign of the $ A \left( 0,-\frac 12\right) $,  $ A \left( -1,-\frac
12\right) $ amplitude. In $ g_L = g_{S+P} = g_{+} = 1 $ units, the
respective effective-mass scales are $\Lambda_{S + P} \sim -35
GeV$, $\Lambda_{f_M + f_E} \sim 53 GeV$.  The occurrence of these
two ambiguities is not surprising because these 3 chiral
combinations only contribute to the L-handed b-quark amplitudes as
$m_b \rightarrow 0$. However, associated with the latter $(f_M +
f_E)$ ambiguity, 3 very interesting numerical puzzles arose at the
$0.1 \%$ level in the ``(+)" amplitudes versus the SM's pure
$``(V-A)"$ amplitudes, see Table 1 :

The 1st puzzle is that the $A_{+} (0,-1/2)$ amplitude for $g_L
+g_{f_M+f_E}$ has the same value in $g_L = 1 $ units, as the
$A_{SM} (-1,-1/2)$ amplitude in the SM; see the corresponding two
``220" entries in the top of Table 1. From the empirical t-quark
and W-boson mass values [5], the mass ratio $ y = \frac{m_W} {m_t}
= 0.461 \pm 0.014$. This can be compared with the puzzle's
associated mass relation
\begin{eqnarray*}
1-\sqrt{2}y-y^{2}-\sqrt{2}y^{3}=x^{2}(\frac{2%
}{1-y^{2}}-\sqrt{2}y)-x^{4}(\frac{1-3y^2}{(1-y^2)^3})+\ldots \\
=1.89x^{2}-0.748x^{4}+\ldots
\end{eqnarray*}
which follows by expanding the $A_{+} (0,-1/2)$ amplitude, given
in the appendix, in the mass ratio $x^{2}=(m_{b}/m_{t})^{2}$.
Before expanding, to express the $A_{+} (0,-1/2)$ amplitude in the
mass ratios $x$ and $y$, the relative strength of the additional
``$(f_M + f_E)$" coupling is specified by substituting
$\Lambda_{+} = E_{W} / 2 = \frac{m_t } {4 } [ 1 + y^2 -  x^2]$ in
$g_L = g_{+} =1$ units [ see discussions following (8) and (9)
below ]. Since empirically $x^{2}\simeq 7 \cdot 10^{-4}$, there is
only a 4th significant-figure correction from the finite b-quark
mass to the only real-valued solution $y=0.46006$ ($m_{b}=0$) of
this mass relation. \ Note versus theoretical interpretation of
this mass relation that $x^{2}$ is very small and not of order
one.  [ The $0.1 \%$ level of agreement of the two ``220" entries
of Table 1 is due to the present central value of $ m_t $, and to
the central value and $ 0.05 \%$ precision of $ m_W$.  The error
in the empirical value of the mass ratio $y$ is dominated by the
current $ 3\%$ precision of $ m_t$. ]

The 2nd and 3rd numerical puzzles are the occurrence of the same
magnitudes of the two R-handed b-quark amplitudes $A_{New} =
A_{g_L =1} / \sqrt \Gamma $ for the SM and for the case of $( g_L
+g_{f_M+f_E} )$. This is shown in the $ A \left( 0, \frac
12\right) $ and $ A \left( 1, \frac 12\right) $ columns in the
bottom half of Table 1.   Except for the differing partial width,
$\Gamma_+ = 0.66 GeV$ versus $\Gamma_{SM}= 1.55 GeV$,  by tuning
the magnitude of L-handed amplitude ratio to that of the SM, we
found that the R-handed amplitude's moduli became those of the SM
to the $ 0.1\%$ level.  [As explained  below following (8), for
$\Lambda_{+} = E_W/2$ the magnitudes of these two R-handed moduli
are actually exactly equal and not merely numerically equal to the
$ 0.1\%$ level.]

Notice that due to the additional $f_M + f_E$ coupling, it is the
$\mu =\lambda _{W^{+} } -\lambda _b = - 1/2 $ helicity amplitudes
$A_{New}$ which get an overall sign change.

\section{Four types of analytic relations}

The first three of the four analytical relations, (i)-(iv) below,
are a deeper realization of the just discussed 2nd and 3rd
numerical puzzles: The first type of relation is (i):
\begin{equation}
\frac{A_{i} (0,1/2) } { A_{i} (-1,-1/2) } = \frac{1}{2}
\frac{A_{i} (1,1/2) } { A_{i} (0,-1/2) }
\end{equation}
This holds separately for $i=$(SM), (+).  (i) relates the two
amplitudes which change or ``flip" sign, i.e. the amplitudes with
$\mu=\lambda_W - \lambda_b=-1/2$, to the two amplitudes which do
not in the case of the 3 numerical puzzles. The second type of
relation is (ii):
\begin{equation}
\frac{A_{+} (0,1/2) } { A_{+} (-1,-1/2) } = \frac{A_{SM} (0,1/2) }
{ A_{SM} (-1,-1/2) }
\end{equation}
Notice that this relation relates the $t \rightarrow W^+ b$
amplitudes in the two cases, (SM) and (+).   (ii) is for the
sign-flip amplitudes, c.f. Table 1, but the analogous relation
holds for the non-sign-flip amplitudes. [This non-sign-flip
relation is not independent of (i) and (ii); it follows by simply
substituting (4) with respectively $i=$(+), (SM) on each side of
(5).]

These 3 equations, the two in (4) and one in (5), occur to all
orders in the $y$ and $x$ mass ratios.  (5) and the (+) one in (4)
are also independent of the effective-mass scale $\Lambda_{+}$.
This independence with respect to specific values of $x,y$ and
$\Lambda_{+}$ is an analytic generalization of the numerical
agreement discussed in Section 1. Due to (ii), the two equations
in (i) imply but are stronger than simply an independent equality
of the ratios of the (SM) and (+) amplitudes which do/do-not
change sign, i.e. the existence of 3 equations is also a stronger
result than is apparent from only the 2nd and 3rd numerical
puzzles.

On the other hand, the two equations in (i) can be rewritten to
relate the ratios of left-handed and right-handed amplitudes for
each coupling, that is
\begin{equation}
\frac{A_{i} (0,-1/2) } { A_{i} (-1,-1/2) } = \frac{1}{2}
\frac{A_{i} (1,1/2) } { A_{i} (0,1/2) }
\end{equation}
for $i=$(SM), (+). Consequently, by determining the effective mass
scale $\Lambda_{+} = \Lambda_{+}(m_W/m_t, m_b/m_t)$ so that there
is an exact equality for the ratio of left-handed amplitudes
(iii):
\begin{equation}
\frac{A_{+} (0,-1/2) } { A_{+} (-1,-1/2) } = -
 \frac{A_{SM} (0,-1/2) } { A_{SM} (-1,-1/2) },
\end{equation}
the normalized $A_{New} = A_{g_L =1} / \sqrt \Gamma $ amplitudes
for the SM and for the case of a $( g_L +g_{f_M+f_E} )$ coupling
are all  exactly equal in magnitude to all orders in $y$ and $x$,
with $\Lambda_{+} \sim 53 GeV$. From (7), in $ g_L = g_{+} = 1$
units, the analytic formula for $\Lambda_{+}$ is
\begin{equation}
\Lambda_{+} = \frac{m_t } {4 } [ 1 + (m_W / m_t)^2 -
 (m_b / m_t)^2]
\end{equation}
or equivalently, $\Lambda_{+} = E_W/2$ where $E_W$ is the energy
of the final W-boson in the t-rest frame.  Eq.(7) is the third
type of relation; it requires (8) to hold whereas (i) and (ii) do
not.

Therefore, due to (i) and (ii), relation (iii) implies an S-matrix
probability condition that the $A_{New} = A_{g_L =1} / \sqrt
\Gamma $ amplitudes are {\bf exactly equal} in magnitude between
the SM and (+) case, to all orders in the two mass ratios.  Only
the actual value of this new EW scale $\Lambda_{+} \sim 53 GeV$
depends on the empirical values of $m_W/m_t$, $m_t$, and the fact
that the $m_b/m_t$ ratio is small. This S-matrix ``locking
mechanism" for arbitrary $x$ and $y$, as described by relations
(i) thru (iii), supports the simple interpretation that the 2nd
and 3rd numerical puzzles arise due to a large chiral weak-moment
of the top quark. Although a large chiral weak-moment,
$\Lambda_{+} = E_W/2 \sim 53 GeV$, changes the $t_R$ to $b_L$
transition amplitude, it does not drastically effect the $SU(2)_L$
X $U(1)_Y$ gauge structure.  With present knowledge, it is also
less radical to consider an unexpected intrinsic property of the
top-quark itself, i.e. an anomalous ``moment", instead of a
tree-level occurrence of an additional EW coupling.

The SM is a theory, being renormalizable and unitary.  In such a
framework, an anomalous moment coupling implies other new physics,
e.g. new particle production, at an effective mass scale of $ \sim
2 \Lambda_+ \sim 106 GeV $. Unfortunately in regard to detailed
predictions, such models are rather non-unique in regard to the
particle multiplets and their associated couplings; they are also
complex and complicated in regard to a satisfactory treatment of
higher order effects.  Here the additional R-chirality t-quark
weak coupling could be problematic given (i) the successful
agreements of the SM through the one-loop level versus EW
observables and (ii) the SM's self-consistency of 3 point-like
fermion families; however, due to possible form-factor effects
there need not be any inconsistencies. When there is additional
empirical information on $ t \rightarrow W^{+} b $ and if , for
instance, a deeper exact symmetry were found which yields the
tWb-transformation, see (10) below, this should expedite
development of a theory with a top-quark with a large chiral
weak-moment.

The final type of relation is simply the first numerical puzzle,
i.e. the equality (iv):
\begin{equation}
A_{+} (0,-1/2) = A_{SM} (-1,-1/2),
\end{equation}
It is important to note that (9) assumes that $\Lambda_{+} = E_W
/2$. The numerical agreement of the two sides of (9) and the
empirical agreement of the associated mass relation discussed in
Section 1 depend on the empirical value of the mass ratio
$m_W/m_t$, on having set the relative coupling strength
$\Lambda_{+} = E_W /2$, and on $ m_b / m_t < < m_W / m_t $.

The top-quark mass is the most accurately measured quark mass. It
is the measured values of the $W$-boson and top-quark masses that
make this (iv) relation something that is not understood.  If the
measured value of $m_W/m_t$ were different, (iv) would not hold.
If $\Lambda_{+}$ is not set equal to $ E_W /2 \sim 53 GeV$, (iv)
would not hold. Therefore, relations (i) thru (iii) lead to the
possibility of (iv) and its equivalent realizations holding for
the current values of $ m_W / m_t$ and $m_b / m_t $ with their
present precisions [5].

It is not obvious whether (iv) is an exact or approximate relation
and, unfortunately, the empirical masses will not be better known
for some time.  The equality (iv) is equivalent to
$\sqrt{2}=\frac{q_W}{E_{W}}\left( \frac{E_{W}+q_W}{m_{W}}\right) $
. Thus, (iv) is also equivalent to the velocity formula
$\sqrt{2}=v\gamma (1+v)=v\sqrt{\frac{1+v}{1-v}}$, where $v$ is the
velocity of the W-boson in the t-rest frame, so by this cubic
equation $v=0.6506\ldots$ without input of a specific value for
$m_b$. Other $Q_{em} = - \frac{1}{3}$ quark decay modes are
predicted by the standard model and the observed W-boson velocity
must vary with different flavor $Q_{em} = - \frac{1}{3}$ quark
mass values, so $v=0.6506\ldots$ is presumably modified by higher
order corrections depending on the $Q_{em} = - \frac{1}{3}$ quark
mass.  In the context of these four analytic relations, we think
that the relation of $v$ and $m_{Q_{em} = - \frac{1}{3} }$ is a
very important, fundamental open issue that has become manifest by
this equivalent realization of (iv).  To empirically investigate
(iv), a measurement of $v$ at a $l^- l^+$ collider near the
$t\bar{t}$-threshold might eventually be better than using the
empirical mass ratio $ y = \frac {m_W} {m_t} $.  [ In the $ m_b =
0$ limit, the formula $ v = \sqrt{2} \frac{m_W}{m_t} c $
transforms this cubic equation into the mass relation listed in
Sec. 1 and inversely; in this formula the speed of light has not
been set equal to 1. ]

In summary, the four analytic-type relations can be characterized
by a tWb-transformation $A_{+}=M$ $A_{SM}$  where $M=v$
$diag(1,-1,-1,1)$ \ due to (i) thru (iii), and where
\begin{eqnarray}
A_{SM}=[A(0,-1/2),A(-1,-1/2),A(0,1/2),A(1,1/2)] \nonumber \\
=
\sqrt{m_{t}\left( E_{b}+q_W \right) } \: [\frac{\sqrt{2}} {v},\sqrt{2},- \frac{v} {\sqrt{%
2} }(\frac{m_{b}}{E_{b}+q_W}),-\sqrt{2}(\frac{m_{b}}{E_{b}+q_W})]
\end{eqnarray}
with $v$ the real root of $\sqrt{2}=v\sqrt{\frac{1+v}{1-v}}$ due
to relation (iv).  In (10), (iv) has been used for the two
components involving the longitudinal component of the W. It is
not clear whether there is a dynamical mechanism and/or a
mathematical-symmetry origin for such a tWb-transformation.

\section{Remarks}

Unfortunately, model-dependent interpretations and assumptions are
needed to relate the above analytic realizations to the observed
$t \bar{t}$ production [2]. Experimental tests and measurements at
the Tevatron [2] and LHC [3] should be able to clarify situation,
for instance by excluding simple phenomenological possibilities.
Here in (1) and (2) we discuss two simple models:

(1)  As the simplest assumption, the top-quarks $t$ described by
the SM weak decay amplitude and those $T$ described by the (+)
weak decay amplitude are presumed identical except for their
differing chiral weak-moments.  Since (i) nothing sufficiently
fundamental appears to forbid that the observed top is more than
one top-like object and since (ii) the explanation of the first
numerical puzzle (9) might be dynamical, it is important to
perform simple tests so as to empirically constrain such a
possibility from $t \bar{t}$ hadroproduction. In regard to the
model we now construct, these are our motivations; it is not a
matter of any presently available empirical data on $t \bar{t}$
hadroproduction disagreeing with a SM prediction.  The $t \bar{t}$
cross-section has a $25\%$ precision from Run I at the Tevatron
which is expected to be reduced to $10\%$ in Run IIa [5]. Our
limited objective is to make a simple model to investigate how one
can empirically restrict the possibility that in hadroproduction
the observed top-quark is a linear combination of $t$ and $T$,
specifically $ t_{obs} = c_t t + c_T T$ where $c_{t,T}$ are
unknown complex coefficients.  The observed anti-top-quark is $
\bar{t}_{obs} = {c_t}^* \bar{t} + {c_T}^* \bar{T}$ and the
asterisk denotes complex conjugation. We also assume that
production of $t \bar{t}$ and $T \bar{T}$ combinations occur and
quantum-mechanically interfere in hadroproduction, but that the
production of $t \bar{T}$ and $T \bar{t}$ combinations is
forbidden. Note that for this model to be viable, simultaneous
non-zero values of both $|c_{t,T}|^{2}$ are required.  Given
continued agreement of QCD with more precise hadroproduction data,
the viability of this model will require that neither $t$ nor $T$
couple with full QCD strength. This might occur due to possible
form-factor effects. Alternative and/or more detailed models of
this type could, of course, be constructed.

Tests of this model are possible because the (+) types, $T$, would
occur with a longer lifetime and because there are simple
differences in stage-two spin-correlation functions:  If for
simplicity only $\lambda_b = - \frac{1}{2}$ amplitudes are
considered, then for the general angular distribution in the
$(t\bar{t})_{cm} $ frame, i.e. in Eq.(62) of [6], the product of
decay density matrices $ R_{\lambda _{1}\lambda _{1}^{^{\prime }}}
( t \rightarrow W^+ b \rightarrow \ldots )  \overline{R_{\lambda
_{2}\lambda _{2}^{^{\prime }}}} ( \bar{t} \rightarrow W^- \bar{b}
\rightarrow \ldots ) $ for the top and anti-top is to be replaced
according to
\begin{eqnarray}
R_{\lambda _{1}\lambda _{1}^{^{\prime }}}   \overline{R_{\lambda
_{2}\lambda _{2}^{^{\prime }}}}  & \rightarrow & [\Gamma
_{L}\overline{\Gamma _{L}}+\Gamma
_{T}\overline{\Gamma _{T}}+I\overline{I}\cos \theta \cos \overline{\theta }%
](|c_{t}|^{2}+v^{2}|c_{T}|^{2})^{2} \nonumber \\
&&+[\Gamma _{L}\overline{\Gamma _{T}}+\Gamma _{T}\overline{\Gamma _{L}}%
](|c_{t}|^{2}-v^{2}|c_{T}|^{2})^{2} \nonumber \\ &&+[I\cos \theta
(\overline{\Gamma _{L}}+\overline{\Gamma _{T}})+(\Gamma
_{L}+\Gamma _{T})\overline{I}\cos \overline{\theta }%
](|c_{t}|^{4}-v^{4}|c_{T}|^{4})
\end{eqnarray}
This equation is a compact, schematic display of the
terms arising from the, now four, composite decay density matrices $%
R_{\lambda _{1}\lambda _{1}^{^{\prime }}} \ldots$ per
\begin{equation}
R_{\lambda _{1}\lambda _{1}^{^{\prime }}}=\Gamma _{\lambda
_{1}\lambda _{1}^{^{\prime }}}+I_{\lambda _{1}\lambda
_{1}^{^{\prime }}}\cos \theta ,\ldots
\end{equation}
where $\Gamma $ does not depend on W longitudinal/transverse
interference ( the helicity parameters $\xi $ and $\sigma =\zeta $
in the case of non-coexistence) and $I$ depends on W
longitudinal/transverse interference ( the $\eta _{L}$ helicity
parameter in the case of non-coexistence). The term ``schematic"
means, for instance, that the angle $\theta$ in (12) represents
the angles/momenta specifying $W^+ b \rightarrow \ldots$ which can
be used for separating the $I$ versus $\Gamma$ terms in
$R_{\lambda _{1}\lambda _{1}^{^{\prime }}}$, c.f. [6]. The
helicity parameters of [6] were normalized by the partial width
assuming a single kind of t-quark whereas in (11) the overall
normalization includes a factor $
(|c_{t}|^{2}+v^{2}|c_{T}|^{2})^{2} $.  From (11), measurement of
both this factor and, e.g., the ratio $
(|c_{t}|^{2}-v^{2}|c_{T}|^{2})^{2}  /
(|c_{t}|^{2}+v^{2}|c_{T}|^{2})^{2} $ can be used to constrain the
possibility of simultaneous non-zero values of $|c_{t,T}|^{2}$. By
(11), it is evident that a W longitudinal/interference measurement
is not needed, i.e. the $\Gamma_i \overline{\Gamma_j}$ terms are
sufficient.  If indirect signatures for such a model were to be
found, it would obviously then be desirable to perform a direct
lifetime experiment to confirm and better investigate the
empirical phenomena.  It is to be emphasized that (11) is model
dependent regarding non-$W^+ b$-channel interactions of the $t,T$.
 In single top-production [7] by W-gluon fusion, the
unknown $|c_{t,T}|^{2}$ factors do not occur, and both $t$ and $T$
would be produced by their respective weak current couplings, per
(1) and (2).

(2)  Fortunately, in the simpler non-coexistence case where either
$t$ or $T$ is the observed top-quark, a sufficiently precise
measurement of the sign of the $\eta_L \equiv \frac 1\Gamma
|A(-1,-\frac 12)||A(0,- \frac 12)|\cos \beta _L $ helicity
parameter will determine the sign of $cos \beta_L $ where $
\beta_L = \phi _{-1}^L- \phi _0^L $ is the relative phase of the
two $\lambda_b = - \frac{1}{2} $-amplitudes, $A(\lambda
_{W^{+}},\lambda _b)=|A|\exp (i\phi _{\lambda _{W^{+}}}^{L,R})$.
Measurement of the sign of $
 \eta_L =  \pm 0.46$(SM/+) due to the large interference
between the W longitudinal/transverse amplitudes could exclude
such a large chiral weak-moment.  Second, measurement of the
closely associated $ {\eta_L}^{'} \equiv \frac 1\Gamma
|A(-1,-\frac 12)||A(0,- \frac 12)|\sin \beta _L $ helicity
parameter would provide useful complementary information, c.f.
second paper in [1]. In the absence of $T_{FS}$-violation,
${\eta_L}^{'} =0$. The possibility of $CP$ violation effects in
top-quark physics is reviewed in [8]. In the case of the SM's
top-quark, next-to-leading order QCD corrections to top-quark spin
correlations at hadron colliders have been recently calculated in
[9].

By single top-quark production [7] at a hadron collider the
partial width for $t \rightarrow W^+ b$ can be measured and so the
$v^2$ factor-difference in their associated partial widths can
also be used to determine whether the observed top-quark is the
SM's $t$ ( with $\Gamma_{SM}= 1.55 GeV$ ), or a $T$ ( with
$\Gamma_+ = 0.66 GeV$). More generally, single top-production at
hadron and at linear colliders can provide important information
on the Lorentz-structure of the $tWb$-vertex [10,11]. For hadron
colliders, the effects of a finite top-quark width in a variety of
top-quark production processes have been investigated in [12].

(3)  The observed top-quark may, indeed, turn out as predicted by
the standard model.  In this case, the empirical significance of
the four types of analytic relations is not clear. They might
dynamically arise due to a weak-interaction phase-interface, or
some other mixed-component quantum phenomena between top-quarks
$t$ without the large chiral weak-moment and others $T$ with such
a moment. Such a possibility is indirectly indicated by the fact
that all the analytic relations involve both the (SM) and (+)
amplitudes and by the appearance in the above analytic
realizations of the kinematic variables $E_W$ (as the $ 2
\Lambda_{+}$ value) and $v$, the W-boson velocity in the top
rest-frame (e.g. in the tWb-transformation). For the (+)
amplitudes, the Lorentz structure of the effective coupling is
\begin{equation}
\gamma ^\mu P_L + \iota \sigma ^{\mu \nu } v_\nu P_R
\end{equation}
where $P_{L,R} = \frac{1}{2} ( 1 \mp \gamma_5 ) $ and $v_{\nu}$ is
the W-boson's relativistic four-velocity. From this perspective,
besides higher precision accelerator-experiment mass measurements,
theoretical investigations of mechanisms to produce anomalous weak
and/or electromagnetic interaction moments in quantum field
theoretic systems and in quantum statistical mechanical systems
can be constructive. Such studies could be useful towards (i)
unraveling the physical significance of these analytical
relations, (ii) indicating relevant non-accelerator and/or
accelerator tests, and (iii) relating the mass generation of the
top-quark, the W-boson, and the different flavor $Q_{em} = -
\frac{1}{3}$ quarks.

This work was partially supported by U.S. Dept. of Energy Contract
No. DE-FG 02-86ER40291.

{\bf Appendix:  Helicity amplitudes for $t \rightarrow W^+ b$ }

In terms of all the coupling constants in (1) and (2), from [6]
the helicity amplitudes defined by (3) and in the Jacob-Wick phase
convention are:

For both $(V\mp A)$ couplings and for $\lambda _b =-\frac
12$,%
\begin{eqnarray*}
 A\left( 0,-\frac 12\right) & = & g_L \frac{E_W +q_W }{m_W }
\sqrt{m_t \left( E_b +q_W \right) } -g_R \frac{E_W -q_W }{m_W }
\sqrt{m_t \left( E_b - q_W \right) } \\
 A\left( -1,-\frac
12\right) & = & g_L \sqrt{2m_t \left( E_b +q_W \right) } -
g_R\sqrt{2m_t \left( E_b -q_W \right) }.
\end{eqnarray*}
and for $\lambda _b =\frac 12$,%
\begin{eqnarray*}
 A\left( 0,\frac 12\right) & = & -g_L \frac{E_W -q_W }{m_W }
\sqrt{m_t \left( E_b - q_W \right) }  +g_R \frac{E_W +q_W }{m_W }
\sqrt{m_t \left( E_b +q_W \right) } \\
A\left( 1,\frac 12\right) &
= & -g_L \sqrt{2m_t \left( E_b -q_W \right) } +g_R\sqrt{2m_t
\left( E_b +q_W \right) }
\end{eqnarray*}
where $g_L = g_V - g_A$, $g_R = g_V + g_A $.

For $(S \pm P)$ couplings, $g_{S \pm P} = g_S \pm g_P$ the
additional contributions are
\begin{eqnarray*}
 A(0,-\frac 12) & =g_{S+P}( \frac{m_t }{2\Lambda
})\frac{2q_W }{m_W }\sqrt{m_t (E_b +q_W )} +g_{S-P}(\frac{m_t
}{2\Lambda })\frac{2q_W
}{m_W }%
\sqrt{m_t (E_b -q_W )}, \quad A(-1,-\frac 12) & =0
\end{eqnarray*}
\begin{eqnarray*}
 A(0,\frac 12) & =g_{S+P}( \frac{m_t }{2\Lambda
})\frac{2q_W }{m_W }\sqrt{m_t (E_b -q_W )}  +g_{S-P}(\frac{m_t
}{2\Lambda })\frac{2q_W
}{m_W }%
\sqrt{m_t (E_b +q_W )}, \quad A(1,\frac 12) & =0
\end{eqnarray*}

The two types of tensorial couplings, $g_\pm = f_M \pm f_E$ and
$\tilde{g}_{\pm}=g_{T^+} \pm g_{T_5^+}$,  give the additional
contributions
\begin{eqnarray*}
A\left( 0,\mp\frac 12\right)         & = & \mp g_{+} (
\frac{m_t}{2\Lambda }) \left[ \frac{E_W \mp q_W }{m_W} \sqrt{m_t
\left( E_b \pm q_W \right) } - \frac{m_b }{m_t} \frac{E_W \mp q_W
}{m_W } \sqrt{m_t \left( E_b \mp q_W \right) } \right] \\
                                             &   & \pm g_{-}
( \frac{m_t }{2\Lambda }) \left[
 - \frac{m_b }{m_t }
\frac{E_W \pm q_W }{m_W } \sqrt{m_t \left( E_b \pm q_W \right) } +
\frac{E_W \pm q_W }{m_W } \sqrt{m_t \left( E_b \mp q_W \right) }
\right] \\
                                           &   & \mp \tilde
g_{+} ( \frac{m_t }{2\Lambda }) \left[ \frac{E_W \pm q_W }{m_W }
\sqrt{m_t \left( E_b \pm q_W \right) } + \frac{m_b }{m_t }
\frac{E_W \mp q_W }{m_W } \sqrt{m_t \left( E_b \mp q_W \right) }
\right] \\
                                             &   & \pm \tilde
g_{-} ( \frac{m_t }{2\Lambda }) \left[
  \frac{m_b }{m_t }
\frac{E_W \pm q_W }{m_W } \sqrt{m_t \left( E_b \pm q_W \right) } +
\frac{E_W \mp q_W }{m_W } \sqrt{m_t \left( E_b \mp q_W \right) }
\right]
\end{eqnarray*}
\begin{eqnarray*}
A\left( \mp 1,\mp\frac 12\right) & = & \mp \sqrt{2} g_{+} (
\frac{m_t }{2\Lambda }) \left[
  \sqrt{m_t \left( E_b
\pm q_W \right) } -  \frac{m_b }{m_t }
 \sqrt{m_t \left( E_b \mp
q_W \right) } \right] \\
                                             &   & \pm
\sqrt{2} g_{-} ( \frac{m_t }{2\Lambda }) \left[
 - \frac{m_b }{m_t }
 \sqrt{m_t \left( E_b \pm
q_W \right) } +  \sqrt{m_t \left( E_b \mp q_W \right) } \right]
\\
                                  &   & \mp \sqrt{2} \tilde
g_{+} ( \frac{m_t }{2\Lambda }) \left[
 \sqrt{m_t \left( E_b
\pm q_W \right) } + \frac{m_b }{m_t }
 \sqrt{m_t \left( E_b \mp
q_W \right) } \right] \\
                                             &   & \pm
\sqrt{2} \tilde g_{-} ( \frac{m_t }{2\Lambda }) \left[
  \frac{m_b }{m_t }
 \sqrt{m_t \left( E_b \pm
q_W \right) } +  \sqrt{m_t \left( E_b \mp q_W \right) } \right]
\end{eqnarray*}

The present paper uses the Standard Model's (SM) and the $(f_M +
f_E)$'s amplitudes:  In $g_L = g_{+} = 1$ units and suppressing a
common overall factor of $\sqrt{m_t \left( E_b +q_W \right) }$,
from the above expressions these helicity amplitudes are with
$y=m_W / m_t$:

For only the $(V-A)$ coupling
\begin{eqnarray*}
 A_{SM} \left(
0,-\frac 12\right) & = & \frac{1 }{y } \; \frac{E_W +q_W }{m_t }
\\
 A_{SM} \left(
-1,-\frac 12\right) & = & \sqrt{2}  \\
 A_{SM} \left( 0,\frac 12\right)
& = & -  \frac{1 }{y }  \frac{E_W -q_W }{m_t } \left( \frac
{m_b}{m_t-E_W +  q_W}  \right) \\
 A_{SM} \left( 1,\frac 12\right) & = & -
\sqrt{2} \left( \frac {m_b}{m_t-E_W +  q_W}  \right)
\end{eqnarray*}
and for only the $(f_M + f_E)$ coupling
\begin{eqnarray*}
 A_{f_M + f_E} \left(
0,-\frac 12\right) & = &  - ( \frac{m_t }{2\Lambda_+ }) \;  y  \\
 A_{f_M + f_E} \left(
-1,-\frac 12\right) & = & - ( \frac{m_t }{2\Lambda_+ }) \sqrt{2}
\; \frac{E_W +q_W }{m_t } \\
 A_{f_M + f_E} \left( 0,\frac 12\right)
& = & ( \frac{m_t }{2\Lambda_+ }) y \left( \frac {m_b}{m_t-E_W +
q_W}  \right) \\
 A_{f_M + f_E} \left( 1,\frac 12\right) & = & ( \frac{m_t }{2\Lambda_+ })
\sqrt{2} \;  \frac{E_W - q_W }{m_t } \left( \frac {m_b}{m_t-E_W +
 q_W}  \right)
\end{eqnarray*}

From these, the ``$(V-A) + (f_M + f_E)$" coupling's amplitudes are
obtained by \newline $A_{+} (\lambda_W, \lambda_b) = A_{SM}
(\lambda_W, \lambda_b) + A_{f_M + f_E } (\lambda_W, \lambda_b)$.

\begin{center}
{\bf Table Captions}
\end{center}

Table 1:  Numerical values of the helicity amplitudes $ A\left(
\lambda_{W^{+} } ,\lambda_b \right) $ for the standard model and
for the 2 dynamical phase-type ambiguities (with respect to the
SM's dominant $\lambda_b = - 1/2$ amplitudes).  The values are
listed first in $ g_L = g_{S+P} = g_{+} = 1 $ units, and second as
$ A_{new} = A_{g_L = 1} / \surd \Gamma $. The latter removes the
effect of the differing partial width in the $(f_M + f_E)$ case.
The respective effective-mass scales for these 2 dynamical
phase-type ambiguities are $\Lambda_{S + P} \sim -35 GeV$,
$\Lambda_{f_M + f_E} \sim 53 GeV$. [$m_t=175GeV, \; m_W =
80.35GeV, \; m_b = 4.5GeV$ ].


\begin{thebibliography}{333}

\bibitem{adler} C.A. Nelson, in Proceedings of IVth Rencontres du
Vietnam, ``Physics at Extreme Energies," (July 13-25, 2000,
Hanoi); C.A. Nelson and L.J. Adler, Jr., Eur. Phys. J. {\bf C17},
399(2000); and see C.A. Nelson and A.M. Cohen, Eur. Phys. J. {\bf
C8}, 393(1999).  In these papers on tests for `` $(V-A)$ $+$
single additional Lorentz structures" in $t \rightarrow W^+ b$
decay, the term ``dynamical phase-type ambiguity" was introduced
to refer to the 2 cases in which the sign change in the
$\lambda_b= - \frac{1}{2}$ amplitudes arises from the single
additional Lorentz structure. This is to be contrasted to the
mathematical forcing of a ``phase-ambiguity" by simply changing
by-hand the sign of one or more of the four helicity amplitudes
$A(\lambda_W, \lambda_b)$.
\bibitem{fnal} CDF collaboration, T. Affolder, et.al., Phys.Rev.Lett. {\bf 84},
216(2000); D{\O} collaboration, B. Abbott, et.al., Phys.Rev.Lett.
{\bf 85}, 256(2000).
\bibitem{2} ATLAS Technical Proposal, CERN/LHCC/94-43,
LHCC/P2 (1994); CMS Technical Design Report, CERN-LHCC- 97-32;
CMS-TDR-3 (1997).
\bibitem{nlc} M. Pohl, hep-ex/0007039; http://tesla.desy.de/ ;
Japanese Linear Collider Group, JLC-I, KEK-Report 92-16(1992);
American Linear Collider Working Group Report, J. Bagger, et.al.
hep-ex/0007022; Proceedings of 1999 Int'l Linear Collider
Workshop, Sitges, eds. E. Fernandez and A. Pacheco (U.A. Barcelona
Pubs.); P.D. Grannis, 2000 Linear Collider Workshop Summary,
hep-ex/0101001.
\bibitem{5}  $m_W =80.434 \pm 0.037 GeV$ from LEP and $p\bar{p}$
experiments, G. Bella at TAU2000 workshop; $m_t = 174.3 \pm 5.1
GeV$, PDG2000; and the pole mass $m_b \sim 4.6 \pm 0.2$, e.g. see
talks at ICHEP2000.  Expected precisions for $m_t$ and $m_W$
measurements from future experiments were reported at the 2001
Intl. Europhysics Conference on High Energy Physics by D.
Charlton, G. Chiarelli, and D. Cavalli and for the Tevatron's Run
II at the 20th Lepton-photon Symposium by Y.-K. Kim:  $\delta m_t
= \pm 3 GeV$ per experiment from run IIa at the Tevatron, and
$(\delta m_t, \delta m_W) = (\pm 1.5 GeV, \pm 0.030 GeV)$ per
experiment from runs IIa and IIb combined; $(\pm 1.5 GeV, \pm
0.015 GeV)$ per experiment from the LHC, and $(\pm [ 0.1-0.2 ]
GeV, \pm 0.006 GeV)$ from a NLC.
\bibitem{nklm1} C.A. Nelson, B.T. Kress, M. Lopes, and T.P. McCauley, Phys. Rev.
{\bf D56}, 5928(1997).
\bibitem{st1} S. Willenbrock and D. A. Dicus, Phys. Rev. {\bf D34},
155(1986); C.-P. Yuan, {\it ibid.}{\bf D41}, 42 (1990); R.K. Ellis
and S. Parke, {\it ibid.}{\bf D46}, 3785(1992); G. Bordes and B.
van Eijk, Z. Phys. {\bf C57}, 81(1993); and T. Stelzer, Z.
Sullivan, and S. Willenbrock, Phys. Rev. {\bf D56}, 5919(1997).
\bibitem{CP} D. Atwood, S. Bar-Shalom, G. Eilam, and A. Soni, Phys. Rept.
{\bf 347}, 1(2001).
\bibitem{SCqcd} W. Bernreuther, A. Brandenburg, Z.G. Si and P.
Uwer, hep-ph/0111346; Phys.Rev.Lett. {\bf 87}, 242002-1(2001);
Phys. Lett. {\bf B509}, 53(2001); and see references therein. For
QCD corrections at $l^- l^+$ colliders, see C. Macesanu,
hep-ph/0112142 and its references.
\bibitem{st2} E. Boos, L. Dudko and T. Ohl, Eur. Phys. J.
{\bf C11}, 473(1999); A.Belyaev and E. Boos, Phys. Rev. {\bf D63}
034012(2001); and E. Boos, M. Dubinin, M. Sachwitz, and H.J.
Schreiber, Eur. Phys. J.{\bf C21}, 81(2001).
\bibitem{st3} T.M.P. Tait and C.-P. Yuan, Phys. Rev. {\bf D63}, 014018(2000).
\bibitem{twidth} N. Kauer and D. Zeppenfeld, hep-ph/0107181.


\end{thebibliography}
\end{document}